\documentclass[fleqn,10pt]{wlscirep}
\usepackage[utf8]{inputenc}
\usepackage[version=4]{mhchem}
\usepackage[T1]{fontenc}
\usepackage{subcaption}
\usepackage{siunitx}
\usepackage{svg}
\usepackage{nccmath}
\usepackage{multirow}
\usepackage{tabularray}
\usepackage{verbatim}
\title{First Experimental Results on Range Monitoring in Carbon Therapy with the TIARA Detector}

\author[1,**]{Maxime Pinson}
\author[1]{Ad\'elie Andr\'e}
\author[2]{Yannick Boursier}
\author[2]{Mathieu Dupont}
\author[1]{Marie-Laure Gallin Martel}
\author[2]{Alicia Garnier}
\author[1]{Christophe Hoarau}
\author[1]{Pavel Kavrigin}
\author[3]{Daniel Maneval}
\author[2]{Christian Morel}
\author[1]{Jean-Fran\c{c}ois Muraz}
\author[4]{Marco Pullia}
\author[4]{Simone Savazzi}
\author[1,*]{Sara Marcatili}

\affil[1]{Université Grenoble Alpes, CNRS, Grenoble INP, LPSC-IN2P3 UMR 5821, 38000 Grenoble, France}
\affil[2]{Aix-Marseille Univ, CNRS/IN2P3, CPPM, Marseille, France}
\affil[3]{Centre Antoine Lacassagne (CAL), 06200 Nice, France}
\affil[4]{Centro Nazionale di Adroterapia Oncologica (CNAO), Pavia, Italy}

\affil[*]{sara.marcatili@lpsc.in2p3.fr}
\affil[**]{maxime.pinson@lpsc.in2p3.fr}


\begin{abstract}
In the context of range monitoring for particle therapy, this study presents the first experimental results obtained with the TIARA detector using carbon-ion beams at the CNAO clinical center in Pavia, Italy. TIARA is based on the Prompt Gamma Timing (PGT) technique, which measures the time of flight (TOF) between incident ions and prompt gamma rays (PGs) emitted during nuclear interactions in the target. While the TIARA prototype has previously been validated with protons, carbons present a more challenging scenario due to their higher linear energy transfer, nuclear fragmentation products, and the continuous beam time structure of synchrotron accelerators. Experiments were performed by irradiating PMMA targets of different thicknesses with 200~MeV/u carbon beams. A coincidence time resolution of 284~$\pm$~32~ps FWHM was achieved, outperforming results previously obtained with protons at the same facility. A range accuracy of 4.74$\pm$0.36~mm at a 2$\sigma$ confidence level was measured at clinical intensity, when considering 5600 detected PGs, corresponding to the grouping of four irradiation spots of 2.4$\cdot$10$^6$ ions each. 
Overall, the results demonstrate that PGT-based range monitoring remains viable for carbon-ion beams, although increased background from secondary charged fragments indicates that detector configuration adaptations are required.
\end{abstract}
\begin{document}

\flushbottom
\maketitle

\thispagestyle{empty}

\section*{Introduction}
First proposed by Wilson in 1946 \cite{wilson_radiological_1946}, proton therapy has emerged in recent decades as a promising alternative to conventional photon therapy. Use of heavier particles such as neon and carbon ions was also studied and first tested in the second half of the last century \cite{tobias_pretherapeutic_1973}. Since then, more than 100 particle therapy centers have been established around the world \cite{mohan_review_2022}. Particle therapy exploits the characteristic depth-dose profiles of accelerated hadrons, which exhibit a high energy deposit towards the end of their range (Bragg peak). These peculiar profiles allow a higher conformity of the dose to the tumor than X-ray radiotherapy \cite{mohan_review_2022}, although some challenges remain to be addressed. In particular, the sensitivity to different sources of uncertainties, such as those associated to 
 patient positioning \cite{liebl_influence_2014}, conversion of Hounsfield Units to relative stopping powers \cite{espana_impact_2010}, or anatomical changes between dose fractions \cite{paganetti_range_2012}, make it difficult to harness the full potential of particle therapy.\\
A promising mitigation strategy could emerge from the accurate control of range uncertainties during treatment. To this end, many efforts have been made to measure the range of protons or ions using secondary particles produced by the nuclear interactions occurring in the patient during treatment \cite{kraan_range_2015, krimmer_prompt-gamma_2018, parodi2018}. 
Focusing on carbon-ion therapy, several approaches have been proposed for range monitoring. These include positron emission tomography (PET) to image the activity of $\beta$$^+$ emitters induced by nuclear fragmentation \cite{pennazio_carbon_2018}, the tracking of secondary proton fragments to reconstruct their trajectories \cite{patera, henriquet_interaction_2012}, and, the Prompt Gamma Imaging (PGI) technique \cite{richter_first_2016, idrissi_first_2024, testa_monitoring_2008} exploiting the correlation between the spatial distribution of PG emission vertices and the dose \cite{min_prompt_2006}. \\
PGs are of particular interest for range monitoring. Thanks to their high energy (approximately between 1 and 10 MeV) \cite{kozlovsky_nuclear_2002}, they can exit the patient with minimal interaction, thus carrying largely unbiased information about the nuclear processes at their origin. Moreover, since their emission occurs within a few picoseconds following the nuclear reaction, their spatial correlation with the dose also implies a temporal correlation. 
On this basis, the Prompt Gamma Timing (PGT) technique \cite{hueso-gonzalez_first_2015, marcatili_ultra-fast_2020, heller_demonstration_2024} relies on measuring the combined time of flight of the primary ion and the emitted PG to indirectly monitor the ion range.
In recent years, our collaboration has developed a detection system specifically designed for PGT, referred to as the Time-of-Flight Imaging Array (TIARA). The system combines a fast beam monitoring device \cite{andre_fast_2025}, operated in time coincidence with an array of gamma-ray detectors \cite{Andre_gamma} surrounding the anatomical region of interest. The current prototype comprises eight gamma-ray detection modules, while the full clinical-scale implementation is foreseen to include up to thirty modules.
TIARA has been extensively validated with proton beams \cite{marcatili_ultra-fast_2020, jacquet_high_2023, andre_fast_2025, Andre_gamma} both experimentally and through simulations. It was shown that, in order to achieve a range accuracy of a few millimeters at the 2$\sigma$ confidence level, a coincidence time resolution (CTR) of the order of 100 ps RMS or 235 ps FWHM, is required.
In this work, we extend its application to carbon-ion beams, with the aim of assessing the experimental feasibility of PGT-based range monitoring in this more complex irradiation scenario.\\
It is not straightforward to predict a priori whether PGT will perform better or worse with carbon ions compared to protons. On the one hand, several properties of carbon-ion beams suggest potential advantages. Their higher specific energy loss improves time resolution in beam monitoring systems, while reduced multiple Coulomb scattering (MCS) leads to less straggling in the ion TOF, and consequently improved CTR. In addition, the lower instantaneous intensities typically associated to carbon treatments reduce pile-up in the beam monitor, allowing individual ions to be tagged even at clinical intensities. This is a notable limitation of PGT with protons, particularly in cyclotron based accelerators, where many particles arrive simultaneously in short bunches and only the bunch arrival time can be measured, resulting in poor CTR often exceeding 1~ns RMS.\\
On the other hand, carbon-ion fragmentation introduces a significant challenge.
Lighter projectile fragments extend beyond the Bragg peak, generating a fragmentation tail that broadens the TOF distribution and reduces the sharpness of its distal fall-off, particularly due to secondary charged particles and PGs reaching the detector.
In this work we investigate the feasibility of PGT at the CNAO hadrontherapy center, addressing the broader question of its applicability with synchrotron beams. At such facilities, during the extraction spill, the beam is effectively continuous on the timescale of gamma detectors, so multiple PGs may arrive simultaneously rather than in well-separated bunches, potentially leading to false coincidences that could degrade TIARA response.\\
This paper aims to address these key aspects and is structured in three main sections, dedicated respectively to evaluating the feasibility of PGT with synchrotron beams, quantifying the time resolution of the TIARA modules, and assessing the range accuracy achievable with carbon ions. In a fourth section, the feasibility of downstream detection of PGs is also discussed.
\subsection*{The TIARA detector} 
\begin{figure}[h!]
    \centering
    \subfloat[\centering]{{\includegraphics[width=.425\linewidth]{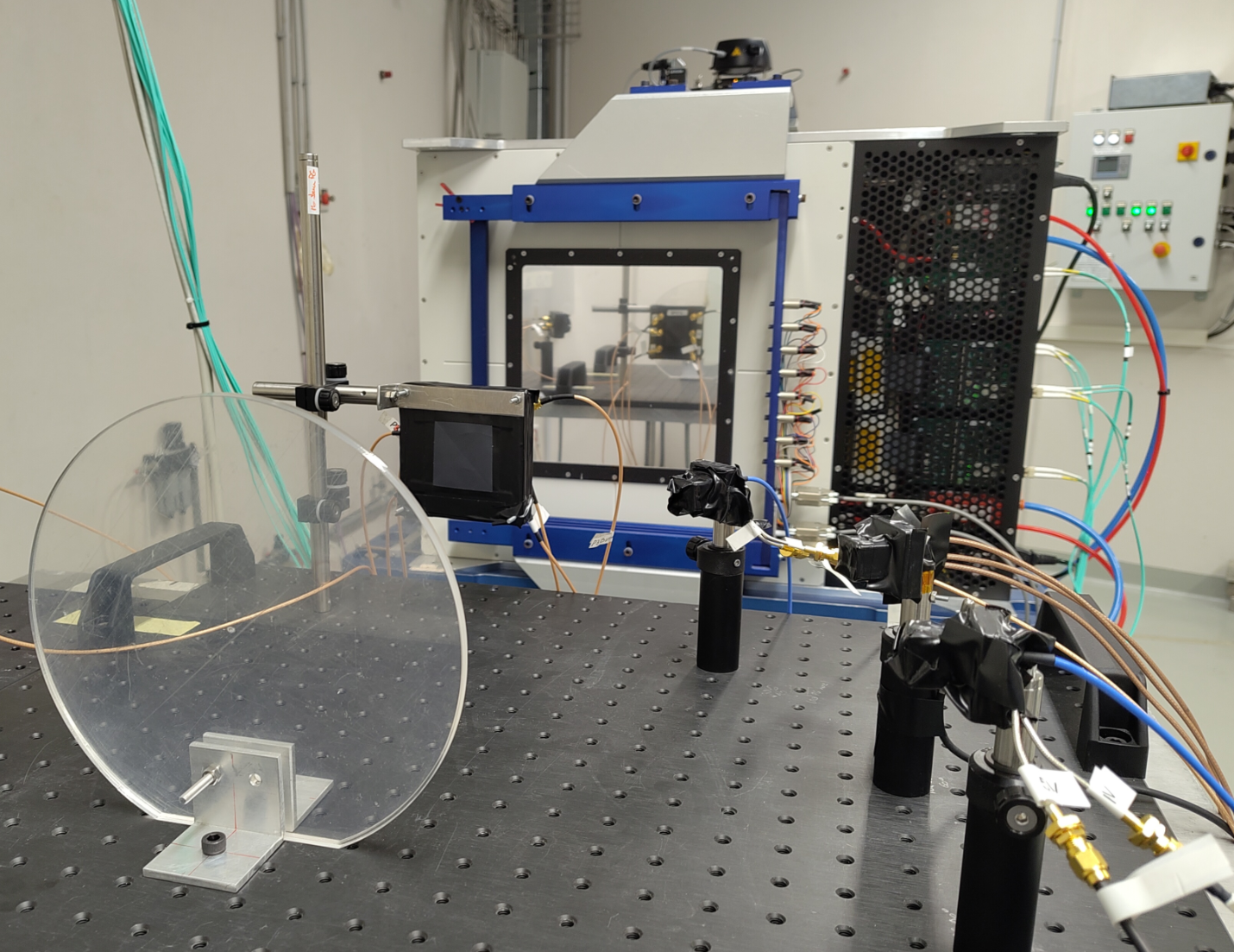} }}%
    \qquad
    \subfloat[\centering]{{\includegraphics[width=.5\linewidth]{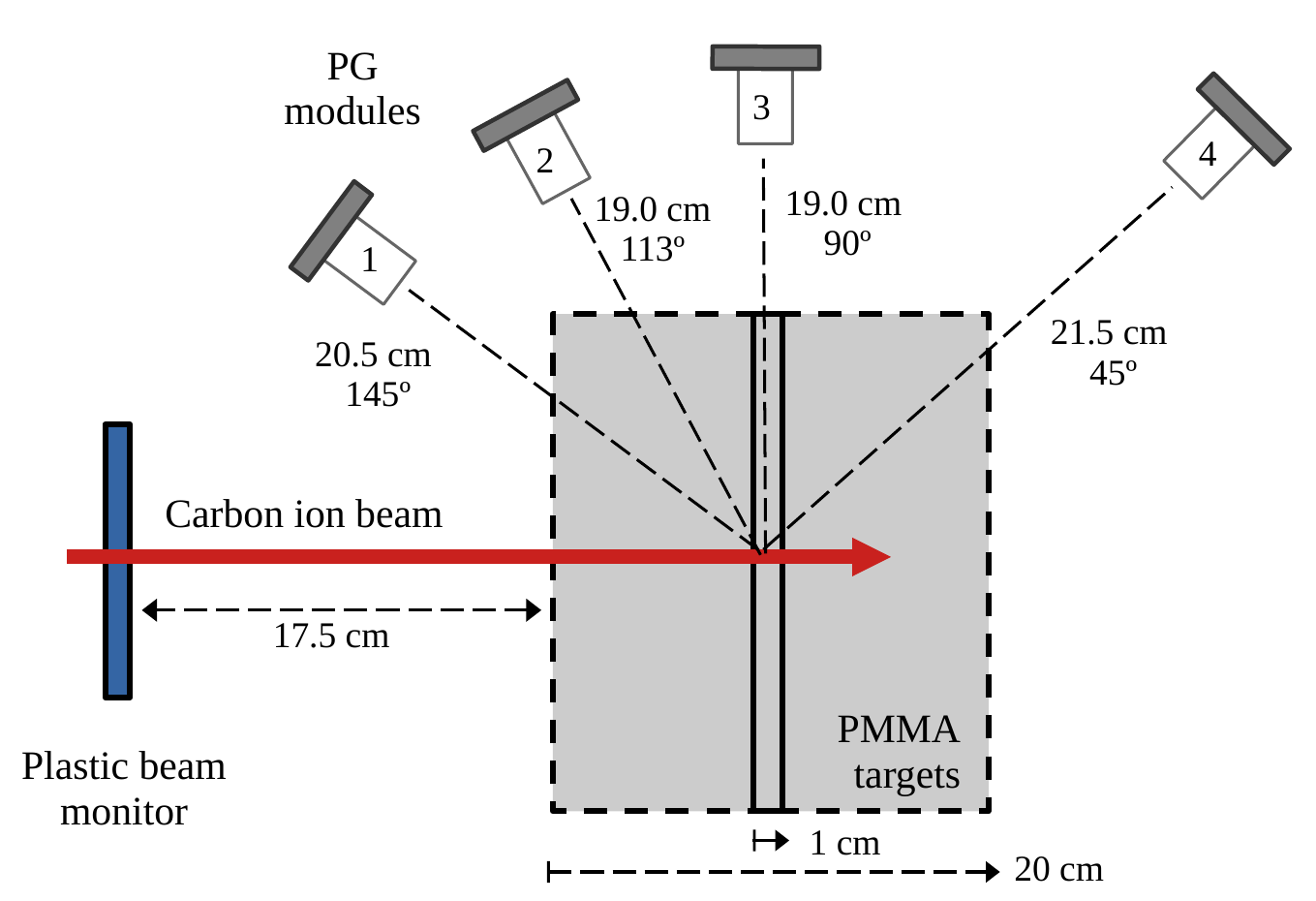} }}%
    \caption{\textbf{(a)} Photo of the TIARA prototype setup at CNAO, showing the thin PMMA calibration target, the beam monitor, and three PG modules. \textbf{(b)} Schematic of the experimental setup: the beam (in red) goes through a 0.5 mm plastic beam monitor, followed by a PMMA target (either 1~cm or 20~cm thick, depending on the experiment). A beam dump is also present  $\sim$1.5~m downstream of the beam monitor. Three gamma modules are placed upstream of the target at approximately 20~cm from the beam axis, while one is placed downstream. }%
    \label{fig:thin_target_schema_photo}%
\end{figure}

TIARA consists of an array of small-volume gamma ray detectors (8 in the current version) arranged around the patient and operated in time coincidence with a fast plastic scintillator beam monitor. Dedicated PG detection modules have been developed, each consisting of a $1.5 \times 1.5 \times 2$ cm$^3$ PbF$_2$ Cherenkov crystal read out by a $2\times2$ array of $6\times6$ mm$^{2}$ SiPMs from Hamamatsu (S13360-6075PE) \cite{Andre_gamma}, corresponding to an active photodetection area of 64\%. The monitor and three PG modules are pictured in Fig.\ref{fig:thin_target_schema_photo}(a).\\
The beam monitor is a modified version of the one previously developed for protons~\cite{andre_fast_2025}. It is based on an EJ-204 fast plastic scintillator of $25\times25\times0.5$ mm$^3$, instrumented with sixteen $3\times3 $ mm$^2$ SiPMs (Hamamatsu S13360-3075CS) arranged in 4 strips along the edges of the scintillator.
In order to accommodate for the higher linear energy transfer (LET) of carbon-ion beams (approximately 20 to 30 times greater than that of protons in the relevant energy range), the monitor thickness was halved (from 1 mm to 0.5 mm), and one amplification stage was removed from the front end to avoid the saturation of the acquisition system. \\
TIARA has been extensively characterised with proton beams at different accelerators, resulting in CTRs as low as 252~$\pm$~3~ps FWHM, for 63 MeV protons from a cyclotron at single proton regime (SPR) \cite{Andre_gamma}. In earlier measurements performed at CNAO, a CTR of 349~$\pm$~16~ps FWHM was achieved with 100~MeV protons \cite{Andre_gamma}.
\section*{Results}
\subsection*{Prompt Gamma Timing with a synchrotron accelerator}
The CNAO synchrotron, located in Pavia, Italy, is a clinical facility that delivers protons (60 MeV to 250 MeV) and carbon ions (120 MeV/u to 400 MeV/u) for hadron therapy treatments. In a clinical configuration, the synchrotron delivers a beam with a FWHM between 4 mm and 10 mm, depending on the selected energy, and the beam position can be adjusted at the iso-center in steps of 0.8 mm with a precision of $\pm$ 0.2~mm \cite{rossi_hadron_2022}. Typical treatment intensities are of the order of $2\cdot10^{9}$~protons/s and $4\cdot10^{7}$~ions/s \cite{giordanengo_cnao_2015}.\\
The TIARA prototype has so far been tested in cyclotron and synchro-cyclotron environments, both of which typically deliver particles in bunches of less than 10~ns width. In these conditions, even the fastest beam monitor, with signals duration of the order of the ns, cannot individually tag in time each proton delivered within the bunch. 
Until recently, to prevent pile-up in the beam monitor, operation has therefore been limited to SPR, corresponding to approximately one particle per bunch. However, the beam intensities employed during these tests are several orders of magnitude below clinical values. 
In contrast, synchrotrons feature significantly different beam structures, with wider bunch widths (of the order of $\sim$100 ns or more) and longer bunch periods. In a synchrotron for non-ultra-relativistic beams, in fact, the revolution period depends on the beam energy, and this periodicity can be observed in the slowly extracted beam when the RF system is switched on during extraction. This occurs at CNAO both with RFKO (Radio Frequency Knock-Out), when the beam is bunched, and with betatron extraction, when empty-bucket channelling is used for spill smoothing.
In the case of CNAO, the revolution frequency for carbon ions ranges between 1.8 MHz (555~ns) and 2.8 MHz (357~ns), depending on the beam energy. At a kinetic energy of 200 MeV/u, the nominal revolution frequency is about 2.2 MHz ($\sim$450~ns). With these time characteristics, during slow extraction, the beam appears quasi-continuous on the time scale relevant for the gamma detector.
This results in lower instantaneous currents compared to cyclotrons,  allowing the use of higher intensities while still avoiding pile-up effects in the beam monitor. Moreover, clinical carbon-ion intensities are approximately one order of magnitude lower than those of protons, further mitigating pile-up risks when operating the monitor at the highest count rates.\\
\begin{figure}[h]
    \centering
    \subfloat[\centering]{{\includegraphics[width=.45\linewidth]{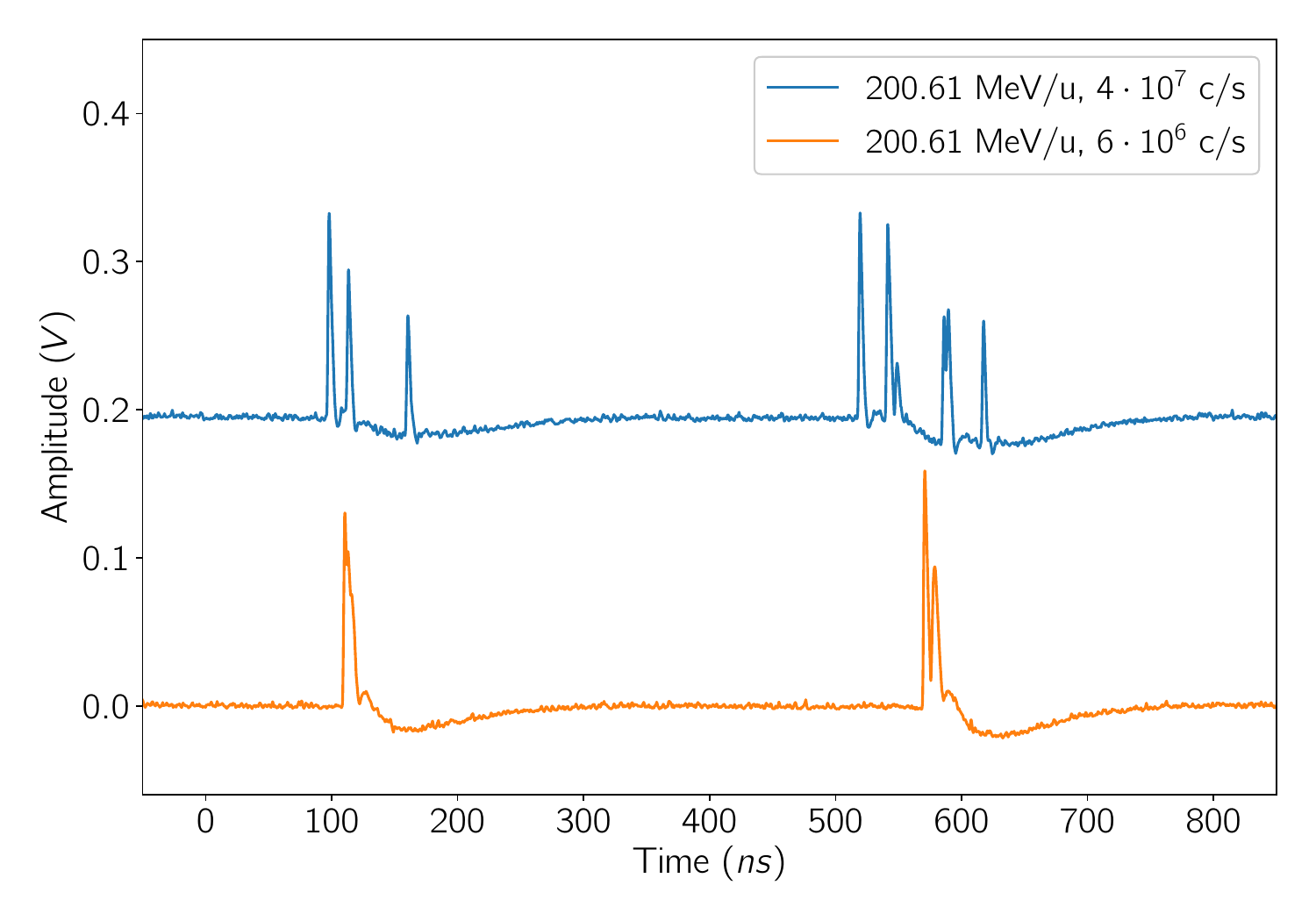} }}%
    \qquad
    \subfloat[\centering]{{\includegraphics[width=.45\linewidth]{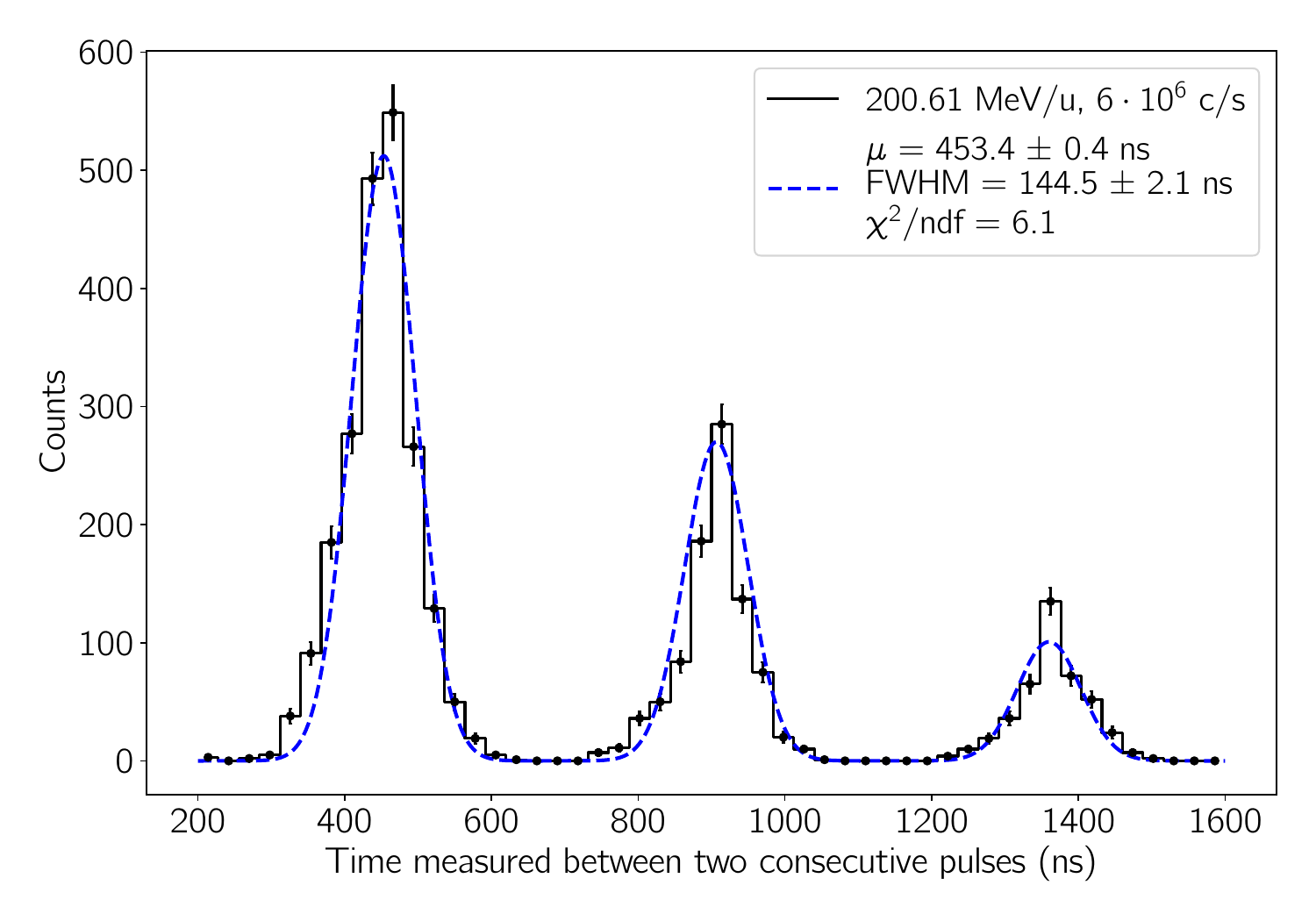} }}%
    \caption{\textbf{(a)} Response of the plastic scintillator beam monitor to 200 MeV/u carbon ions delivered at the CNAO synchrotron at two different intensities: two bunches are visible at this time scale. \textbf{(b)} Histogram of the time between two consecutive pulses obtained for the $6\cdot10^{6}$ ions/s dataset. Error bars represent the statistical uncertainty. The fit is a constrained multi-gaussian with an unique value for the standard deviation, and the centroid values multiples of $\mu$.}%
    \label{fig:marco_and_plastic_intensity}%
\end{figure}
To fully characterise the CNAO beam time structure, the beam monitor was placed along the beam path, as close as possible to the nozzle. Signals from the four readout channels were recorded using a 1 GHz Teledyne LeCroy oscilloscope. The acquisition time window was set to 5 ms in order to capture a large number of ion bunches. Measurements were performed at 200.61 MeV/u for two representative beam intensities: a low intensity of $6\cdot10^{6}$ ions/s, corresponding to the SPR for carbon ions (i.e. Single Ion Regime), and a clinically realistic intensity of $4\cdot10^{7}$ ions/s. Typical results from one of the four channels are shown in Fig.~\ref{fig:marco_and_plastic_intensity}(a). 
Here, two bunches of particles can be identified at a distance of approximately 450~ns. The peaks observed within a single bunch correspond to individual particles in the bunch, and their intra-bunch spacing does not reflect any frequency imposed by the accelerator, but rather the statistical distribution of particles within the bunch population. At both intensities, individual ions are clearly resolved, with only a modest pile-up observed at the higher intensity: an example can be observed in the second bunch of the high intensity data, where a superposition occurs between the second and third protons, as well as between  the fourth and  the  fifth.  All signals, however, display an undershoot, most likely originating from the high-pass response of the monitor’s amplifier, that could limit its use at higher intensities. 
Both pile-up effects and the non-immediate recovery of the signal baseline after ion detection are expected to degrade the accuracy of the timestamp determination.\\
Furthermore, the time interval between consecutive beam pulses was evaluated using large time-scale acquisitions, and the resulting intervals were collected in a histogram for the low-intensity case (Fig.~\ref{fig:marco_and_plastic_intensity}(b)). The histogram shows multiple well-separated peaks with uniform spacing and comparable widths, thus confirming the regularity of bunch separation. The first peak corresponds to intervals between consecutive bunches, while higher-order peaks correspond to separations of two and three bunch periods.
The peak spacing provides a direct measurement of the bunch period, while the peak widths reflect the convolution of the temporal spread of two bunches. 
A constrained fit using Gaussian functions with a common width and mean equal to $\mu \cdot n$ was performed, with $n$ identifying the peak order (1, 2 or 3). 
The multi-Gaussian fit yields a reduced $\chi^2$  of 6.1, indicating that the measured peak shapes are not fully described by an ideal Gaussian model. This discrepancy is most likely due to non-linearities in the detector response, which introduce asymmetric distortions of the peaks. Since the aim of this analysis is to determine the bunch period and the effective bunch width, rather than to model the detailed peak shape, the Gaussian approximation provides sufficiently accurate estimates of these quantities. The extracted values, ($\mu$ = 453.4 $\pm$ 0.4) ns for the bunch period and ($FWHM/\sqrt{2}$ = 102.2 $\pm$ 1.5) ns FWHM after deconvolution, are consistent with the nominal beam parameters.
\subsection*{Coincidence Time Resolution}
The primary objective of this experiment was to evaluate the timing performances of different PG modules.
A 1~cm thick PMMA target placed along the beam axis, served as a point-like source of PGs. A beam dump was also placed  approximately 1.5~m downstream of the beam monitor. 
Three PG modules, positioned at different angles around the target at approximately 20~cm from the beam axis (Fig.~\ref{fig:thin_target_schema_photo}), were tested. The same set of measurements was carried out at 2$\cdot$10$^{6}$ ions/s and at 2$\cdot$10$^{7}$ ions/s.\\
\begin{figure}[!h]
\centering
\includegraphics[width=\linewidth]{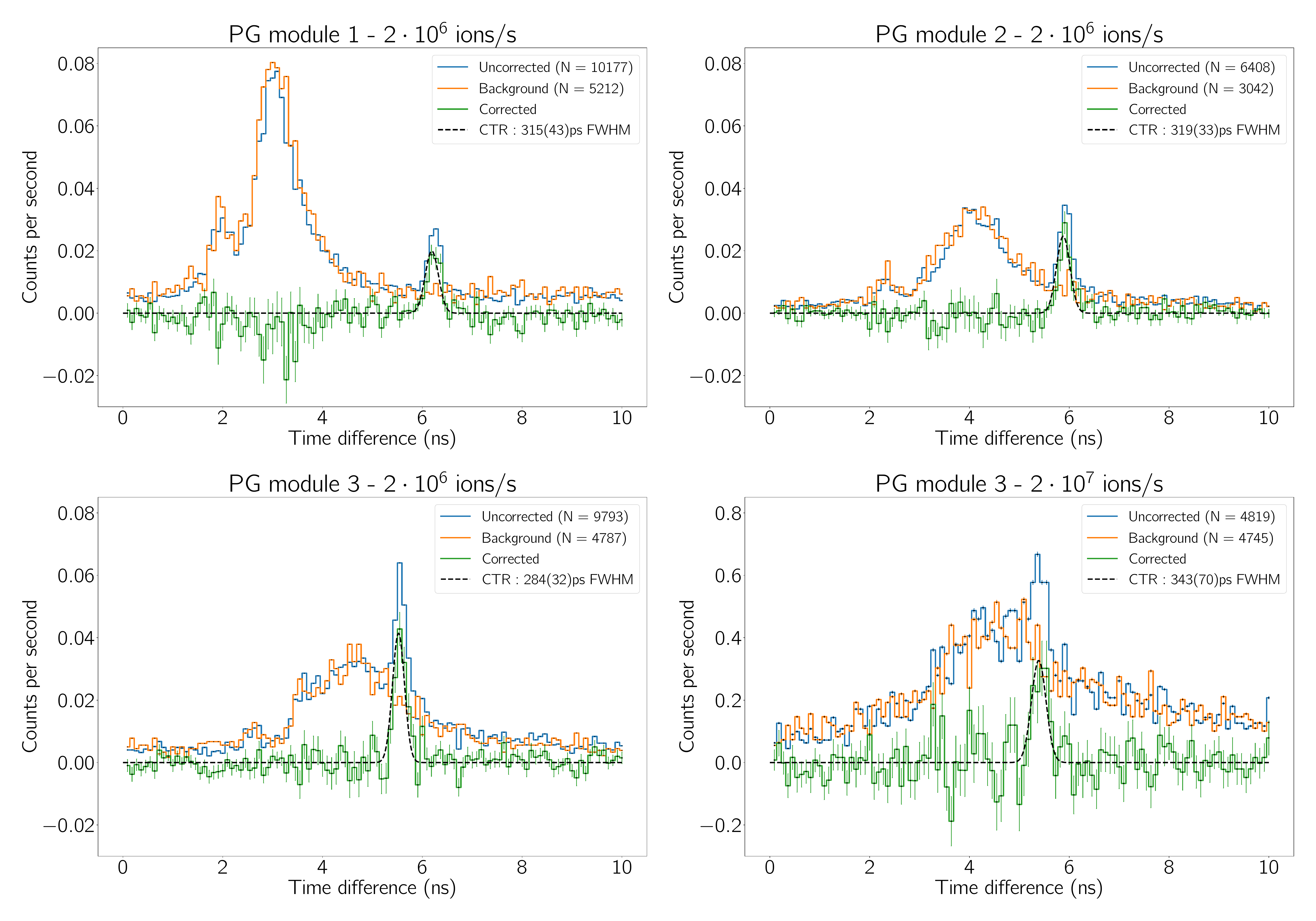}
\caption{TOF distributions obtained from the irradiation of the 1 cm thick PMMA target with a 200 MeV/u carbon beam, normalised to the acquisition time.  Results are shown for the three gamma modules at the intensity of 2$\cdot$10$^{6}$~ions/s (top row, and bottom left), and for module 3 at the intensity of 2$\cdot$10$^{7}$~ions/s (bottom, right). In each plot, the uncorrected datasets correspond to raw data, the background datasets are obtained by removing the PMMA target at acquisition, while the corrected datasets represent the raw data once the background is subtracted. The original number of acquired events ($N$) is disclosed in the legend. For the lower intensity, error bars are within the line size for the uncorrected and background datasets. Corrected data are fit all over the displayed range; reduced $\chi^2$ values are of 1.37, 1.09 and 1.31 for module 1 to 3 at the lower intensity, and of 0.92 for module 3 at the higher intensity. }
\label{fig:calibration_carbon}
\end{figure}
Time stamps from the beam monitor and PG modules were extracted using a digital constant fraction discriminator (CFD) method, and the TOF was calculated as their difference. No corrections were applied for systematic variations in cable lengths. 
The resulting TOF distributions for the low intensity irradiation are displayed in Fig.~\ref{fig:calibration_carbon}. 
To characterise the background in the PG modules, an additional acquisition was performed after removing the PMMA block, allowing contributions unrelated to the target to be identified and subtracted.
Thus, for each PG module, two independent datasets are shown:
one with the thin PMMA target in place, labeled as uncorrected, and one without the target, labeled as Background. 
The corrected distribution, shown in green, is obtained by subtracting the background from the uncorrected measurement, isolating the contribution of PGs originating in the thin PMMA target (see Methods for further details).  A peak around 12~ns corresponding to PGs generated  in the beam dump is also present in the data but was cut-off from Fig.~\ref{fig:calibration_carbon} as it is not relevant for this analysis.   %
The corrected distributions exhibit a Gaussian profile, whose width represents the CTR for the corresponding PG module under the assumption that the straggling of the ion travel time in the target is well below the CTR: for a 1-cm thick PMMA target, this corresponds to 80~ps FWHM.  
The CTRs obtained range from 284~$\pm$~32~ps to 319~$\pm$~33~ps FWHM for the 2 $10^{6}$ ions/s case (Fig.~\ref{fig:calibration_carbon}, top row and bottom left) and are all compatible within the experimental errors. 
For comparison, the CTR previously measured at CNAO for a similar setup with 100~MeV protons in SPR was 349~$\pm$~16~ps~FWHM \cite{Andre_gamma}. The reduced CTR observed with carbon ions is expected, due to their higher energy deposition in the beam monitor, which improves its time resolution \cite{andre_fast_2025}. \\
At the higher intensity (2$\cdot 10^{7}$ ions/s), CTRs remained of the same order of magnitude: 262~$\pm$~76~ps ($\chi^2$/dof = 0.92) for PG module 1 and 509~$\pm$~126~ps ($\chi^2$/dof = 1.02) for PG module 2. 
Fig.~\ref{fig:calibration_carbon} (bottom, right) shows, for comparison, the TOF distributions obtained with PG module 3 where the measured CTR reaches 343~$\pm$~70~ps ($\chi^2$/dof = 0.92). Pile-up and signal undershoot (especially affecting module 2 front-end electronics) at this intensity can occasionally degrade the determination of the detectors' time stamps, affecting the overall timing performance.\\
In contrast to proton beam environments, the background contribution at both intensities is considerably higher for carbon beams due to the additional presence of forward-directed secondary protons and light ions (mainly alphas) from projectile fragmentation reaching the gamma detectors. In Fig.~\ref{fig:calibration_carbon}, the background is pronounced across all three PG modules, particularly for modules positioned close to the beam monitor. Two distinct peaks are evident in the background distribution: a narrow peak at approximately 2-3~ns, depending on the module, and a broader, more prominent peak centered around 3-4~ns. The first peak is attributed to PGs generated within the plastic beam monitor, representing the fastest contribution as its TOF is determined solely by the photon travel time. The second, dominant peak corresponds to secondary charged fragments from the beam monitor (mainly protons and alpha particles) and secondary PGs.  This component is delayed due to the charged particles’ lower velocities, and appears broader because of the wide energy spectrum of fragmentation ions. In general, these contributions are partly reduced for PG modules positioned farther from the beam monitor, as they are also placed farther from the beam line and present a reduced solid angle. It should also be noted that the unfavourable signal to noise ratio (SNR) obtained in this experiment is largely due to the inherently low PG signal resulting from a thin PMMA target. In the next section, we will show that the SNR improves considerably when employing a target that fully stops the carbon beam, thus generating a larger number of PGs. 
\subsection*{Prototype range accuracy}
\begin{figure}[h]
\centering
\includegraphics[width=\linewidth]{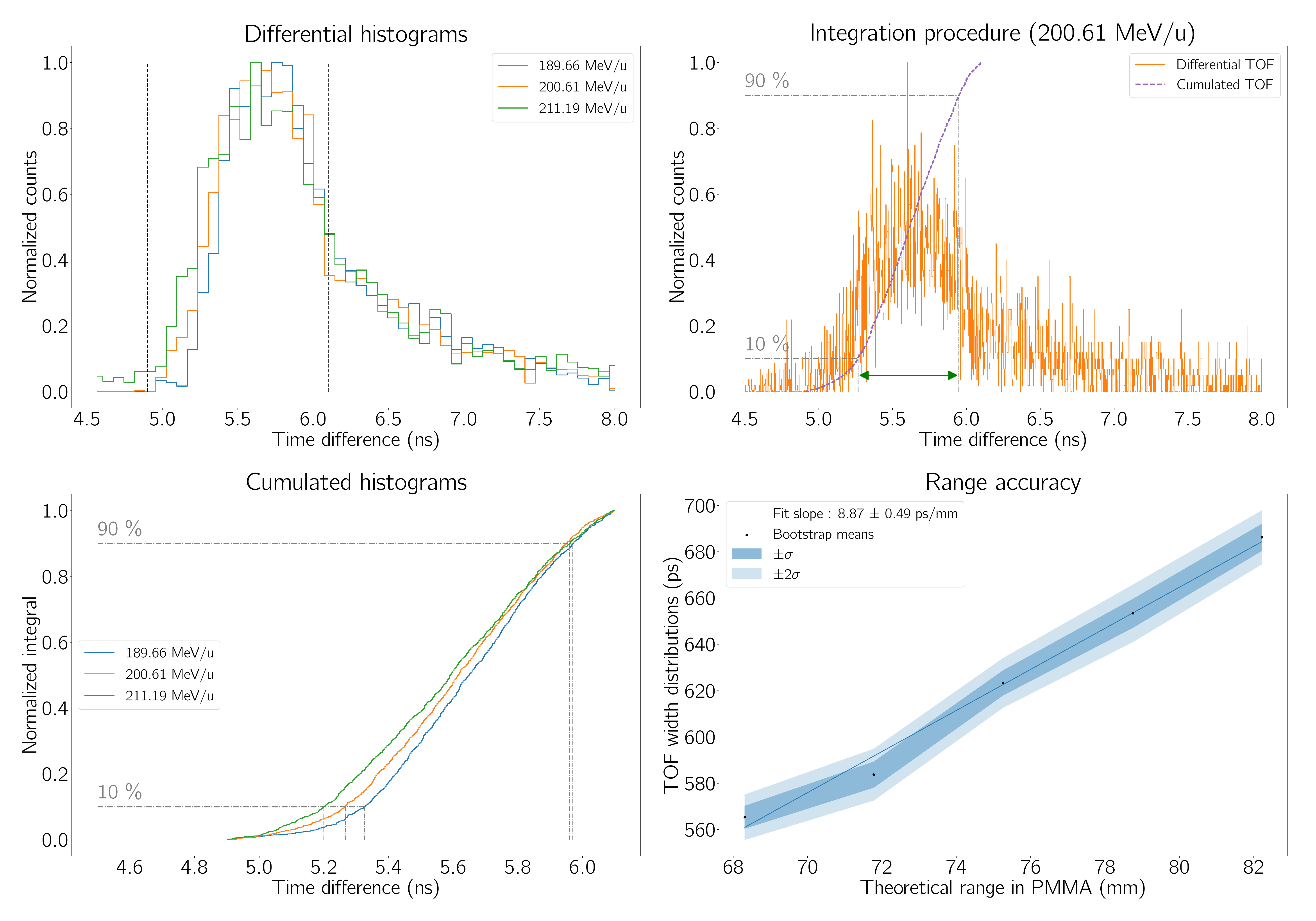}
\caption{ Top, left: background-subtracted TOF distributions for three of the five investigated energies. Histograms are normalised by their maximum and statistical errors are within the line size. The vertical dashed lines show the analysis window. Top, right: the 200.61 MeV/u histogram is rebinned (orange curve) and integrated (purple curve) over the analysis window. Both datasets are normalised to their maximum value. The grey dashed lines indicate the 10th and 90th percentile of the cumulated histogram, while the green  arrow denotes the calculated time width metric. Bottom, left: comparison of the three cumulated histograms obtained from data in the top, left plot. Bottom, right: Results of the bootstrap procedure carried out for a sub-sample size of $N=5600$ for all five energies at $2\cdot10^{6}$ ions/s. The mean TOF widths and their 1$\sigma$ and 2$\sigma$ errors are plotted against the carbon ion range in PMMA.}
\label{fig:bootstrap_allE_and_BS_example}
\end{figure}

To evaluate TIARA performances in a more clinically realistic scenario where the carbon beam is fully stopped within the target, the thin PMMA slab was replaced with a 20~cm-thick block, while the positions of the beam monitor and PG modules remained unchanged. The center  of the thick target was aligned with the center of the thin target, as illustrated in Fig.~\ref{fig:thin_target_schema_photo}(b). \\
\begin{figure}[h]
\centering
\includegraphics[width=\linewidth]{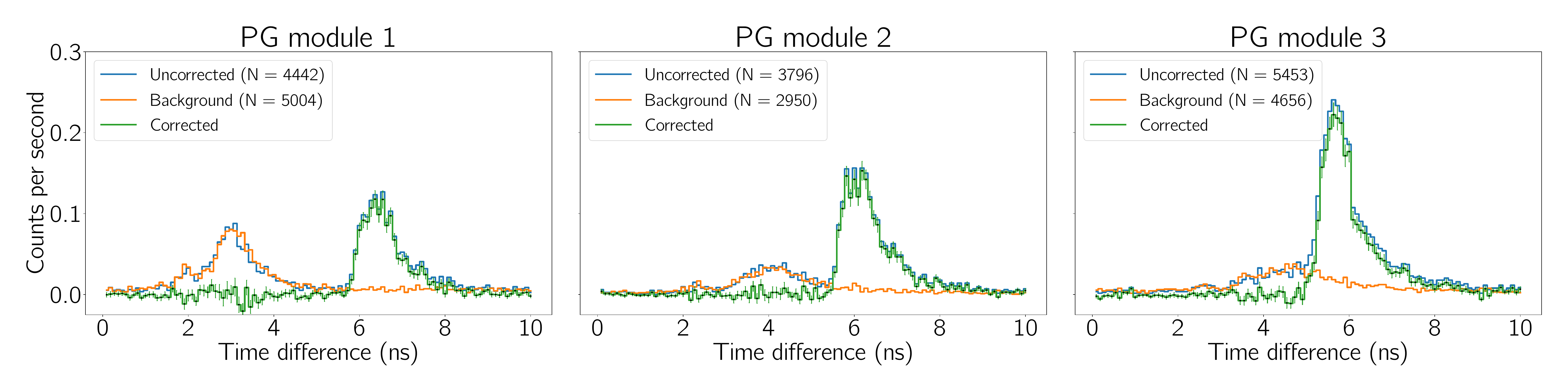}
\caption{Uncorrected TOF distributions, background TOF distributions and corrected TOF distributions for a 200.61 MeV/u carbon beam at 2 $10^{6}$ ions/s impinging on the thick target. The PG signal from the target is centered around 6~ns. In this region, SNRs are 10.5, 14.1 and 8.8, for PG modules 1, 2 and 3 respectively.  Error bars for the uncorrected and background datasets are within the line size.  }
\label{fig:thick_target_carbon}
\end{figure}
Fig.~\ref{fig:thick_target_carbon} shows the TOF distributions measured for a 200.61~MeV/u carbon beam with the three PG modules placed upstream. As for the thin target, background measurements were subtracted from the uncorrected data. While the distribution resulting from PGs and secondary ions from the beam monitor are left unchanged, the signal corresponding to PGs from the target (around 6 ns), is significantly more intense and broader. If only this part of the TOF distribution is considered (i.e. from 5 to 8~ns depending on the chosen module), the SNR becomes relatively large, with values of 10.5, 14.1 and 8.8 for PG modules 1 to 3, respectively. The lower SNR in module 3, in particular, is due to the partial overlap between the monitor and the target signals; in the future, this could be prevented by changing the relative position of the beam monitor and the gamma modules.
In addition, in the background-subtracted distributions, the falling edge exhibits two distinct components: an initial steep decline followed by a more gradual tail. This behaviour is the result of secondary PGs generated from projectile's fragments whose range extends beyond the Bragg peak, and that are undistinguishable from primary PGs at the detector level. \\
In general, the range information of the carbon ions is encoded in the width of the TOF distributions. 
In order to determine TIARA range accuracy, five separate acquisitions were realised with the beam energy varying from 189.66~MeV/u to 211.19~MeV/u. Each energy increment corresponded to a 4~mm range shift in water, that is approximately 3.5~mm in PMMA. This procedure made it possible to 
mimic a potential range shift induced by an inter-fractional change in patient's anatomy. The five irradiations were repeated at two different intensities: 2$\cdot$10$^{6}$~ions/s and 2$\cdot$10$^{7}$~ions/s. \\
The resulting TOF distributions are all similar in shape but differ in their leading edge time positions and widths.
 Fig.~\ref{fig:bootstrap_allE_and_BS_example} top, left shows three of the five measured TOF distributions (the lowest, mid-range and highest energies), after background subtraction for PG module 3. The position of the leading edge is delayed for lower energies, due to the lower ion speed. The total width of the distribution is also affected, since more energetic ions have longer ranges. A measurement of the time shift at the entrance of the target (on the leading edge of the TOF distribution) is not sufficient to compute the ion range, since it would only be a reflection of its speed before reaching the target. A robust metric must instead consider both the rising and falling edges of the distributions to determine their width (in time units).\\
The TOF width is obtained in different steps as  illustrated in Fig. \ref{fig:bootstrap_allE_and_BS_example}. 
First (plot top, left), the analysis is restricted to the primary PG component (before 6.1~ns) as this is the only part of the distribution that actually correlates to the primary ion range. The  analysis window is determined from a  MC simulation (see Methods) and it is represented with two vertical dashed lines. 
Then (plot top, right), each differential TOF histogram (orange curve) is finely binned and integrated (purple curve) within the analysis window to exploit the low pass filtering capabilities of the integral function and thus remove statistical fluctuations. Here, the two datasets corresponding to the 200.61~MeV/u data experiment are both normalised to their maximum for display reasons. 
The bottom, left plot compares the cumulated histograms of three different datasets once normalised by their maximum. The distribution widths in time units are calculated for each dataset as the difference between the 10th and 90th percentile times and reported in the bottom, right plot as a function of the actual carbon range in PMMA for each energy considered. \\
In order to quantify the statistical error associated to the range measurement, i.e. the error bars in Fig.~\ref{fig:bootstrap_allE_and_BS_example} bottom, right, a bootstrap procedure was implemented as detailed in the Methods section.
Indeed, since the TOF width is a global non-linear estimator derived from the full histogram shape, and not a direct observable associated with individual bin counts, propagating independent Poisson uncertainties on the bins is not sufficient to capture the statistical fluctuations of this parameter.
Bootstrap resampling, instead, allows creating lower statistics sub-samples from the original dataset and to perform repeated range  measurements for the same experiment to build the TOF width probability distribution function (pdf). The standard deviation of the resulting TOF width pdf is then interpreted as the statistical uncertainty and reported as 1$\sigma$ and 2$\sigma$ error bars in Fig.~\ref{fig:bootstrap_allE_and_BS_example} bottom, right.
These data provide a calibration between the expected ion range and the one measured in time units: the behaviour is linear within the statistical errors. The 1$\sigma$ range accuracy ($R_{acc, 1\sigma}$) is defined as the minimum distance for which two distinct measurements of range result in a time difference larger than twice the corresponding mean average error of each measurement across the five energies: $2\sigma_{avg}$. In a linear approximation, this can be expressed as a function of the calibration curve slope $A_{fit}$ (Eq. \ref{eq}). %
\begin{ceqn}
\begin{align}
\label{eq}
R_{acc, 1\sigma} = \frac{2 \sigma_{avg}}{A_{fit}}
\end{align}
\end{ceqn}
The uncertainty on the range accuracy is obtained through standard error propagation of Eq. \ref{eq}.\\
In the following, we report the values of $2\sigma$ range accuracy $R_{\mathrm{acc},2\sigma}$ which is simply $2\cdot R_{\mathrm{acc},1\sigma}$.  Range accuracy depends on both the sub-sample size (overall number of detected PGs) and the beam intensity. At CNAO, the nominal beam intensity for carbon-ion treatments is 4$\cdot$10$^{7}$ ions/s, with operational values ranging from 4$\cdot$10$^{6}$~ions/s to 8$\cdot$10$^{7}$~ions/s. In a range monitoring procedure, exploiting the distal irradiation spots is especially relevant as they carry most of the anatomical information and are more sensitive to range shifts caused by inter-fraction variations. A typical irradiation condition at CNAO would therefore consist of a 300~MeV/u carbon beam, delivered in spots that can reach 2.4$\cdot$10$^{6}$ ions. 
Taking into account the detection efficiency expected for the final 30-module version of the TIARA prototype, equal to 0.45\% \cite{Andre_gamma}, and the PG yields for 300~MeV/u carbon ions \cite{krimmer_prompt-gamma_2018}, the expected number of detected PGs per clinical spot can be estimated to be approximately 1400. 
The extrapolation to 30 detectors is considered feasible in the future, as the collaboration is currently finalising a data reconstruction  approach (Prompt Gamma Time Imaging) that allows merging the response of multiple detectors, as well as a calibration procedure  to correct the relative time response of the 30 modules  within an accuracy of few ps.  A preliminary version of the reconstruction algorithm has already been published \cite{jacquet_time--flight-based_2021}.
Clearly,  this approach only provides a first-order estimate of the detected PGs per spot, which ultimately depends on the spot size, beam energy and patient's tissues irradiated. Also, attributing all the events to a single detector is only an approximation, as in the final prototype arrangement not all modules will be placed at exactly 90°. \\
\begin{table}[h]
\begin{center}
\def\arraystretch{1.2}
\begin{tabular}{ |w{c}{2cm}|w{c}{2cm}|w{c}{2cm}|c|c|  }
\cline{4-5}
\multicolumn{2}{c}{} && \multicolumn{2}{c|}{$R_{acc, 2\sigma}$ (mm)}\\
\hline
Clinical spots&Number of ions&$N_{PG}$ (Counts) & $2\cdot10^{6}$ ions/s&$2\cdot10^{7}$ ions/s \\
\hline
\hline
1&$2.4\cdot10^{6}$&1400 & 5.07 $\pm$ 0.30 & 9.30  $\pm$ 0.70 \\
\hline
2&$4.8\cdot10^{6}$&2800 & 3.40 $\pm$ 0.22 & 6.64  $\pm$ 0.50\\
\hline
4&$9.6\cdot10^{6}$&5600 & 2.47 $\pm$ 0.15 & 4.74  $\pm$ 0.36 \\
\hline
\end{tabular}
\end{center}
\caption{\label{tab:sigmas} Extrapolated range accuracies at 2$\sigma$ for three different levels of statistics. The accuracy decreases as the square root of the number of PG events included in the analysis. 1400 events correspond to the number of detected PGs in a distal spot of 2.4$\cdot$10$^{6}$~ions that is expected with the final version of the TIARA detector, including 30 modules (from MC simulation \cite{Andre_gamma}). Results are given for PG module 3.}
\label{tableau}
\end{table}
The bootstrap procedures were performed for both intensities and for three different sub-sample sizes, selected to match the expected statistics for 1, 2 and 4 clinical spots at the CNAO facility. The range accuracies at 2$\sigma$ significance level are reported in Table \ref{tableau} for module 3. The uncertainties on $R_{acc, 2\sigma}$ are calculated by propagating the uncertainty on $A_{fit}$, referenced on Fig. \ref{fig:bootstrap_allE_and_BS_example} bottom, right.
As expected, the range accuracy improves when multiple spots are grouped for range analysis. For example, in the low-intensity configuration, grouping two spots, for an overall statistics of 2800 PGs, yields a range accuracy of 3.40~$\pm$~0.22~mm at 2$\sigma$. This result is consistent with the range accuracy of 3.3~$\pm$~0.1~mm at 2$\sigma$ previously obtained with 63~MeV protons and 3200~PGs \cite{Andre_gamma}. \\
At the higher intensity of 2$\cdot10^{7}$ ions/s, however, the range accuracy is approximately doubled with a value of 4.74 mm when grouping 4 irradiation spots. Nonetheless, if a 1$\sigma$ significance level is accepted, this range accuracy decreases to 2.37~mm.
This degradation is attributed to the reduced precision of the plastic beam monitor in timing the incoming ions, in presence of a moderate pile-up and a non-perfect recovery of the baseline between the detection of two consecutive ions. While pile up in the gamma detector could also occur under extreme conditions, it was not observed at this intensity. \\
Finally, for comparison, the $2\sigma$ range accuracy achieved with modules 1 and 2 is slightly degraded, with values of 3.71~$\pm$~0.23~ps and 3.56~$\pm$~0.15~ps obtained for 4 spots at the lower intensity, respectively. This result is expected, as module 3 preferentially detects PGs originating from the Bragg peak region due to its solid-angle coverage, resulting in improved sensitivity to range variations in this region.
\subsection*{Downstream detection of Prompt Gamma rays}
In a preliminary experiment, a detector module was positioned downstream of the target, at an angle of 45° with respect to the beam axis, with the large PMMA target in place. This configuration was investigated to quantify the contribution of secondary particles from projectile fragmentation and to assess the feasibility of downstream PGT.
In carbon-ion irradiation, nuclear fragmentation produces lighter projectile fragments that emerge with approximately the same velocity as the primary ions. Owing to their lower mass and charge, these fragments exhibit ranges extending beyond the Bragg peak, giving rise to the fragmentation tail. In downstream configurations, such forward-emitted particles may escape the target and reach the detector \cite{robert_distributions_2013,gunzert-marx_secondary_2008}.\\
\begin{figure}[h]
\centering
\includegraphics[width=0.55\linewidth]{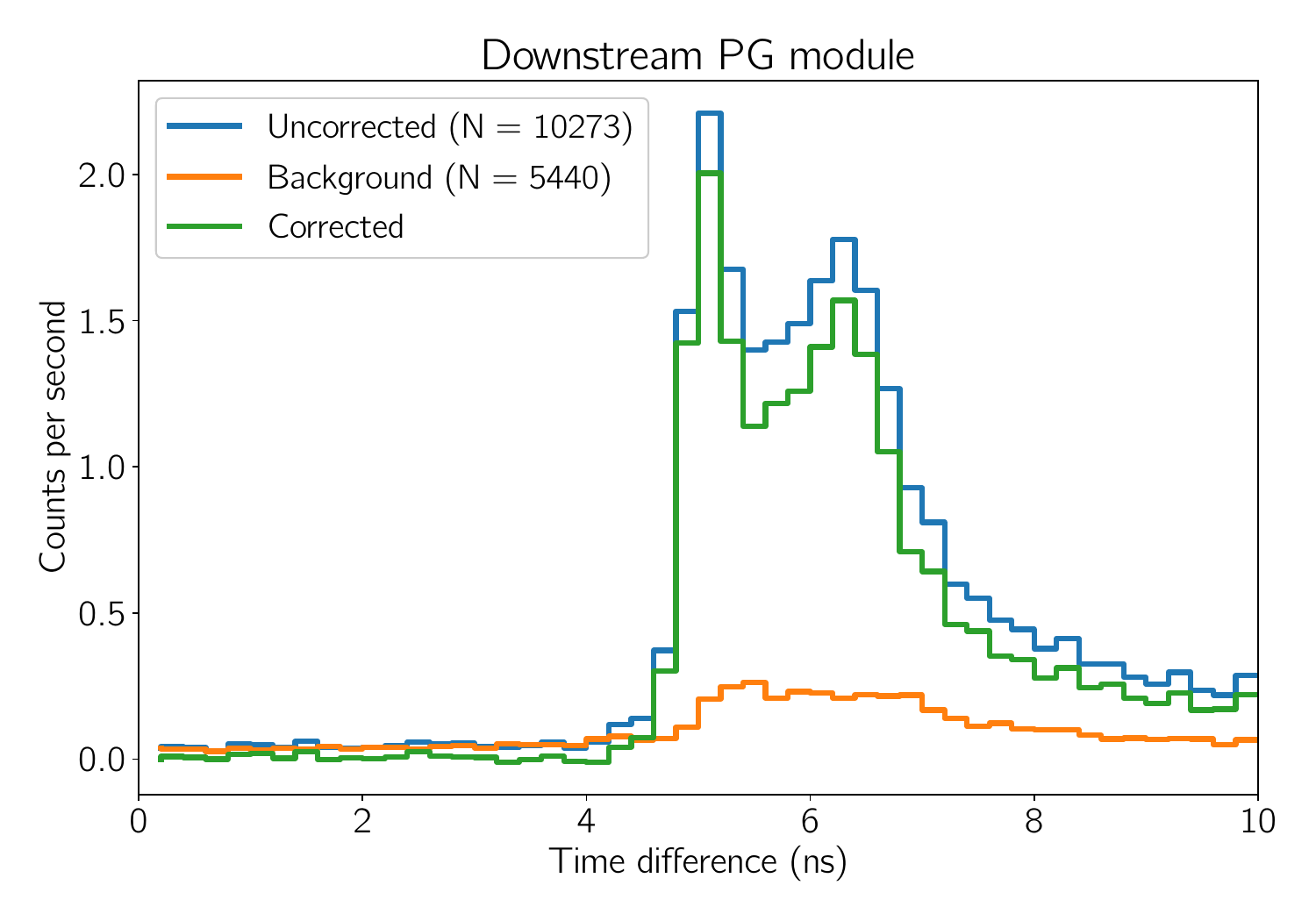}
\caption{TOF distribution measured by a PG detector module placed downstream of the target, at a distance of approximately 20 cm from its center and at an angle of 45° with respect to the beam axis. Two main contributions are observed: the earliest component originates from primary PGs emitted in the target, while the later and more intense contribution is attributed to secondary PGs and forward-directed charged fragments escaping the target. The background spectrum does not include the proton component, confirming that this contribution arises from interactions in the thick target.}
\label{fig:downstream}
\end{figure}
The resulting TOF distributions, shown in Fig.\ref{fig:downstream}, exhibit two distinct components. The first peak, centered at approximately 5~ns, corresponds to PGs emitted by primary carbon ions. The second, broader component arises from the combined contribution of secondary PGs (as those observed in Fig.~\ref{fig:calibration_carbon}), and light ions  directly detected in the module.
Given the wide energy spectrum of the secondary fragments, this contribution leads to a broad TOF distribution that is not correlated with the prompt-gamma emission vertex and therefore with the ion range. Neutrons' contribution is considered negligible as the associated detection efficiency is several orders of magnitude lower than for PGs and protons for TIARA modules \cite{Andre_gamma}.
Based on these findings, downstream detector configurations were considered unsuitable for PGT with the current TIARA setup and were excluded from subsequent measurements.
\section*{Discussion}
The TIARA prototype has demonstrated reliable operation and promising performances with carbon-ion beams at the CNAO synchrotron. While PGT had already been validated with protons at different accelerator facilities, this work addresses the remaining open question of whether the technique can be effectively applied to carbon-ion therapy, despite the substantial differences in beam characteristics and interaction mechanisms.\\
First, this work demonstrates the feasibility of implementing PGT at a synchrotron accelerator. In contrast to cyclotrons and synchro-cyclotrons, which deliver ion bunches on the nanosecond time scale, the CNAO synchrotron operates with a wider bunch duration of approximately 100 ns. At sufficiently low beam intensities (2$\cdot$10$^{6}$ ions/s), individual ions can be resolved by the beam monitor, resulting in optimal coincidence time resolution and range accuracy. At higher intensities (2$\cdot$10$^{7}$ ions/s), close to clinical operation, pile-up effects remain limited due to the wider bunch structure, allowing reliable time stamps to be extracted.\\
A priori, it was not possible to determine whether TIARA would achieve better range accuracy with protons or with carbon ions. While carbon ions exhibit a higher intrinsic PG yield, fewer ions are delivered per spot to achieve the same dose due to their higher LET. 
In the single-ion regime, carbon ions showed a slightly improved performance, both in terms of CTR and range accuracy. At clinical intensities, the CTR is degraded by approximately 26\% while the measured range accuracy is twice as large. This effect can be attributed to multiple factors. A first limitation arises from the current beam monitor design, where signal undershoot prevents a complete return to the baseline between successive ions, even under moderate pile-up conditions. This effect degrades the timing performance but could be mitigated in future developments by optimising the amplifier response. In the same line, further improvements can be expected by reducing the detector thickness to shorten the signal width and increase the detector count rate capability. 
Another critical factor is the ability of correctly identifying carbon-gamma coincidences when multiple ions are detected in a short time window of few ns; the acquisition of false coincidences may produce a time-dependent background below the PG distribution that is difficult to subtract and can change the shape of the TOF profile. \\
More generally, the increased background observed in carbon beams, originating primarily from the beam monitor for upstream modules, is certainly negatively affecting range accuracy. A possible mitigation strategy would be to position the beam monitor closer to the target while keeping the gamma detectors further upstream, thereby reducing the contribution from forward-directed secondary charged fragments and secondary PGs. Nevertheless, the present results demonstrate that range verification with carbon ions at clinical intensities is feasible: by grouping four spots and adopting a 2$\sigma$ (1$\sigma$) significance level, for example, a range accuracy of 4.74~mm (2.37~mm) can be obtained.
Although hadron therapy dose calculations can achieve sub-millimeter numerical precision, clinically relevant in-vivo range uncertainties remain on the order of several millimeters due to CT stopping-power conversion, patient setup error, inter-fractional anatomical changes and motion. Consequently,  safety margins are routinely applied, reaching up to 10 mm for deep-seated tumors \cite{paganetti_range_2012}.
Within this framework, a detector with the range accuracy achieved by TIARA would be capable of detecting clinically relevant discrepancies between the planned and delivered particle range and could therefore contribute to treatment verification.\\
Finally, this work highlights the critical role of detector positioning in achieving optimal range accuracy. In contrast to proton therapy, a downstream measurement configuration does not appear viable with the current detector technology, as secondary fragments significantly degrade the fall-off of the TOF distributions and weaken the correlation between TOF and range. A downstream approach would require either detectors capable of gamma–proton discrimination via pulse-shape analysis or tracking systems able to reconstruct the emission vertex of secondary fragments \cite{patera, henriquet_interaction_2012}. Among the tested configurations, the module placed at 90° with respect to the beam direction provides the best range accuracy, as it is most sensitive to PGs emitted from the Bragg peak region, which carry the strongest range information. While in proton therapy a nearly uniform angular coverage around the patient is envisaged to ensure compatibility with Intensity Modulated Proton Therapy (IMPT), this strategy appears less suitable for carbon-ion therapy. Instead, a ring of detectors placed in a plane orthogonal to the beam direction emerges as a more effective configuration.\\
Overall, these results demonstrate the feasibility and potential of prompt-gamma timing for range verification in carbon-ion therapy. Nonetheless, the present study was performed under simplified experimental conditions; further experiments will be carried out in the next years in order to quantify the range accuracy under more  clinically realistic scenarios, including the use of anthropomorphic phantoms and multiple gamma detection modules.  
\section*{Methods}
\subsection*{Data acquisition and analysis}
Signals from all detectors were acquired using the Wavecatcher digital sampler\cite{breton_wavecatcher_2014} 
with 500 MHz bandwidth and a sampling rate of 3.2 Gs/s. The acquisition was triggered by the coincidence of at least one of the gamma modules with the beam monitor, within a 15 ns time window. The analysis was performed offline.
The time stamps of each signal was extracted by interpolating its rising edge and determining the moment at which the pulse reaches 30\% of its maximum, after baseline correction. The TOF is calculated as the difference between the monitor's and the gamma module's time stamps. \\
Depending on the intensity, due to the typical time structure of the CNAO beam, multiple pulses may be present in the beam monitor on the timescale of a single PG event.
Two strategies were evaluated to manage carbon–gamma coincidence identification at higher intensities.
The first method assigns the TOF using the beam monitor pulse closest in time to the PG signal, after correcting for the systematic delay introduced by cable lengths, whereas the second method computes the TOF for all beam monitor pulses within the same bunch.
The former may reject valid coincidences, while the latter preserves the true coincidence at the cost of reduced signal-to-noise ratio. The first approach was ultimately selected for the PG timing measurements reported in this work.
\subsection*{Background subtraction}
For each TOF histogram acquired (uncorrected), the corresponding background histogram is also measured by removing the main PMMA target. This procedure allows removing the contribution from PGs and secondary charged particles issued by the beam monitor. 
Uncorrected and background histograms are first normalised by the acquisition time and then subtracted to obtain the corrected TOF distributions.  The original number of acquired events ($N$) is disclosed in the legend of relevant figures. Errors are considered Poissonian in the uncorrected and background data, and the variance on the corrected histogram bins is computed as the squared sum of these errors. Negative values resulting from the subtraction are retained.
\subsection*{Determination of the analysis window}
Ion range estimation from TOF histograms requires selecting the portion of the spectrum unaffected by secondary fragments and PGs that are not time-correlated with the primary ion range. This corresponds to the earliest component of the TOF distribution, located before the inflection point at approximately 6.1~ns in Fig.~\ref{fig:bootstrap_allE_and_BS_example}.
To accurately determine this point and define the integration window, a Monte Carlo simulation of the experiment was performed. The irradiation at 200.61~MeV/u was used as the reference configuration, corresponding in a prospective clinical application to the treatment planning conditions.\\
The simulation was carried out using the GEANT4 MC toolkit\cite{agostinelli_geant4simulation_2003} version 10.4.p03, with electromagnetic option 4 and the Liège Intranuclear Cascade (INCL) model\cite{mancusi_extension_2014} for nuclear interactions. The INCL model was selected for its improved agreement with experimental measurements of angular differential cross sections for secondary charged fragments\cite{dudouet_benchmarking_2014} and prompt gamma yields\cite{vanstalle_benchmarking_2017} compared to the Binary Ion Cascade (BIC) and Quantum Molecular Dynamics (QMD) models.\\
The particle transport simulation was coupled to a previously developed response model for TIARA PG modules\cite{Andre_gamma}. The model includes particle- and energy-dependent detection efficiencies derived from optical MC simulations accounting for Cherenkov light production, photon transport, and SiPM matrix detection.\\
This framework enabled the generation of a high-statistics TOF distribution equivalent to that shown in Fig.~\ref{fig:bootstrap_allE_and_BS_example} (top left), but unaffected by statistical fluctuations. The inflection point was numerically determined through polynomial interpolation and used to define the end of the analysis window. The beginning of the window was similarly identified from the inflection point corresponding to the onset of the TOF distribution.
The analysis window determined for the 200.61~MeV/u experiment was subsequently applied to all datasets.

\subsection*{Bootstrap procedure}
A bootstrap procedure was employed to quantify the variability of the TOF width measurement as a function of the number of events $N$ included in the TOF distribution. Each background-corrected TOF distribution (with the negative bins set to zero)
was used as the pdf from which sub-samples were randomly drawn. The TOF width was then computed repeatedly for each sub-sample, as described in figure~\ref{fig:bootstrap_allE_and_BS_example}, to build the TOF width pdf and obtain the corresponding standard deviation value.
To ensure statistical independence of the sub-samples given the finite available statistics, the bootstrap analysis was first performed for 10 sub-sample sizes uniformly spaced between 100 and 800 events. For each sub-sample size, the standard deviation of the TOF width was evaluated. The dependence of the standard deviation on the sub-sample size was subsequently fitted with an inverse square-root function, to extrapolate standard deviations of samples with more than 800 events. 
Then, the bootstrap procedure was applied to samples of 1400, 2800 and 5600 events. The resulting standard deviations are in agreement with those obtain by extrapolation, thus showing that the use of statistical dependent subsamples does not prevent the determination of sound range errors.
\section*{Acknowledgments}
All authors declare that they have no known conflicts of interest in terms of competing financial interests or personal relationships that could have an influence or are relevant to the work reported in this paper. This work was supported by the European Union (ERC project PGTI, grant number 101040381). Views and opinions expressed are however those of the authors only and do not necessarily reflect those of the European Union or the European Research Council Executive Agency. Neither the European Union nor the granting authority can be held responsible for them. This work was partially supported by the European Union’s Horizon 2020 research and innovation programme HITRIplus (grant agreement no. 101008548) via the Transnational Access (TNA) framework.
\section*{Author contributions}

M.Pi. and A.A. analysed the results;
S.M., C.M., Y.B. and M.L.G.M. supervised the analysis;
M.Pi., A.A., S.M., M.D., A.G., D.M. and P.K. conceived the experiments;
M.Pi., A.A., S.M., C.H., M.Pu. and S.S. conducted the experiments;
J.F.M. conceived the detector integration;
C.H. developed the electronics;
M.Pi. and S.M. wrote the manuscript. All authors reviewed the manuscript.

\section*{Data availability statement}

The datasets used and/or analysed during the current study are available from the corresponding author upon reasonable request.

\bibliography{bib}

\end{document}